\documentclass[twocolumn,showpacs,preprintnumbers,amsmath,amssymb]{revtex4}
\usepackage{tabularx,graphicx}\begin{document}
\newcommand{\beq}{\begin{equation}}
\newcommand{\eeq}{\end{equation}}
\newcommand{\beqn}{\begin{eqnarray}}
\newcommand{\eeqn}{\end{eqnarray}}
\newcommand{\bmath}{\begin{subequations}}
\newcommand{\emath}{\end{subequations}}
\title{Spin currents in superconductors}
\author{J. E. Hirsch }
\address{Department of Physics, University of California, San Diego\\
La Jolla, CA 92093-0319}
 
\date{\today} 
\begin{abstract} 
It is argued that experiments on rotating superconductors provide evidence for the existence of macroscopic spin currents in
superconductors in the absence of applied external fields. Furthermore it is shown that the model of hole
superconductivity predicts the existence of such currents in all superconductors. In addition it is pointed out that   spin currents are  required within a related
macroscopic (London-like) electrodynamic description of superconductors recently proposed. The spin current arises through an intrinsic spin Hall effect when negative charge
is expelled from the interior of the metal upon the transition to the superconducting state.
\end{abstract}
\pacs{}
\maketitle 

There is currently great interest in the study of the role of the electron spin degree of freedom in transport in 
solids\cite{spintronics}. In particular the possibility of generating a pure spin current in the absence of charge currents and
magnetization has been 
suggested\cite{spinhall,spinhall2,bhat}, and it was recently reported that it may have been achieved by optical techniques\cite{driel}.

Another area of   recent interest is the
possible existence of novel states of matter that involve spontaneous currents, induced by electronic correlations between
electrons. Early work on superconductivity predicted the existence
of such macroscopic charge currents\cite{heisenberg}, but this was proven impossible by a so-called 'Bloch's theorem'\cite{bohm}. More recently,  
states with local charge currents 
that spontaneously break time reversal invariance have been proposed
to describe high temperature cuprate superconductors\cite{chak,varma}. 

Combining the above two concepts, we proposed several years ago that 
the low temperature phase of certain metals and in particular Chromium may be a novel state of matter induced by electronic
correlations ('spin-split state')  where
parity is spontaneously broken but time reversal invariance is preserved\cite{ss}, in which the system  carries a  macroscopic
spin current but no charge current\cite{ss2}. Such spin currents give rise to electric fields which should be experimentally detectable\cite{ss2,loss}.
Recently Wu and Zhang discussed a variety of interesting possible
states of metals arising from spontaneous generation of spin-orbit coupling\cite{zhang},
one of which is equivalent to the  proposed spin-split state.

As pointed out in our discussion of spin-split states\cite{ss2}, such  spin currents should be insensitive to degradation by non-magnetic
disorder (hence we termed them 'spin supercurrents') for the same reason as supercurrents in superconductors\cite{anderson}: in the presence
of weak disorder, quasiparticles giving rise to the spin current state should be interpreted as time-reversed extended eigenstates of the disordered system
rather than states with
definite crystal momentum k. The question then naturally arises whether spin currents also occur in superconductors.

Note that {\it a single Cooper pair $(k\uparrow, -k\downarrow)$ carries a spin current but not a charge current}, and that the form of the BCS wave function
\beq
|\Psi>=\prod_k(u_k+v_k c_{k\uparrow}^\dagger c_{-k\downarrow}^\dagger)|0>
\eeq
naturally allows for the existence of a spin current if $(u_k,v_k)\neq (u_{-k},v_{-k})$. It is then natural to ask whether such
symmetry breaking will lower the free energy of the system. In this paper we show that while this is not the case for the
attractive Hubbard model, and likely also not the case for electron-phonon models of superconductivity, it is the case when superconductivity
is driven by a kinetic-energy-lowering interaction as in the model of hole superconductivity\cite{hole1}, which has been proposed to describe
all superconducting materials\cite{hole2}. In the model of hole superconductivity, pairing is driven by an off-diagonal matrix element of the
Coulomb interaction; another off-diagonal matrix element of the Coulomb interaction was shown to favor the spin-split
state\cite{ss}. A qualitative picture of the situation envisaged is shown in Figure 1.

\begin{figure}
\resizebox{5.5cm}{!}{\includegraphics[width=7cm]{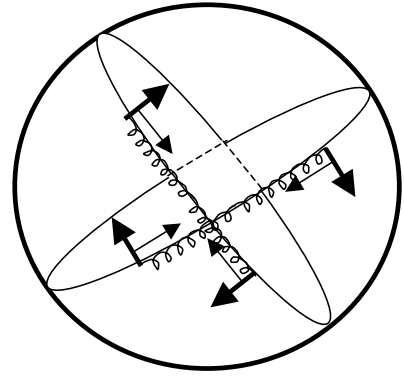}}
\caption{Qualitative picture of Cooper pairs in a spherical superconductor giving rise to spin currents. The arrow perpendicular to the orbit indicates the
direction of the electron magnetic moment, and the arrow parallel to the orbit indicates the direction of motion.}
\label{atom4}
\end{figure}

Before entering into the microscopic modeling we present a macroscopic argument. Consider a rotating simply connected
superconducting body. A uniform magnetic field ('London field') develops in its interior\cite{becker}, in the absence of any applied
external field, given by
\beq
\vec{B}=-\frac{2m_ec}{e}\vec{\omega}
\eeq
with $m_e$ and $e$ the free electron mass and charge respectively, and $\vec{\omega}$ the angular velocity of rotation. This has been
verified experimentally for both conventional\cite{rotating1} and high $T_c$\cite{rotating2} superconductors, and can be 'explained' theoretically\cite{london} from
the London equation
\beq
\vec{J}=-\frac{n_se^2}{m_ec}\vec{A}
\eeq
using
\bmath
\beq
\vec{J}=en_s\vec{v}_s
\eeq
\beq
\vec{v}_s=\vec{\omega}\times\vec{r}
\eeq
\emath
and $\vec{B}=\nabla\times \vec{A}$, with $n_s$ the superfluid density. Eq. (4b) says that in the interior of the superconductor the superfluid
rotates with the same velocity ($\vec{v}_s$) as the body. Instead, within a London penetration depth of the surface the superfluid
'lags' and this relative motion between it and the body provides the electric current that generates the magnetic field Eq. (2)\cite{becker}.

However a superfluid electron rotating with angular velocity $\vec{\omega}$ requires a centripetal force
\beq
\vec{F}_c(\vec{\omega})=-\frac{m_ev_s^2}{r}\hat{r}=\frac{e}{2c}\vec{v}_s\times\vec{B}
\eeq
which is only $half$ of the Lorentz force provided by the magnetic field $\vec{B}$ in Eq. (2)\cite{clarify}. 
Hence this situation requires  a compensating $outward$ electric
force   on the electron to balance the radial force, which would require a {\it{higher density of negative charge in the interior}}
of the superconductor than near the surface. There is no logical reason why such a inhomogeneous charge density would develop in
a rotating superconductor, in particular for why the electrons in a rotating normal metal would suddenly move $in$ when the metal
is cooled into the superconducting state.

Instead we propose the following solution of this conundrum. Assume the speed of electrons of spin $\sigma$ in the
rotating superfluid is given by
\beq
\vec{v}_\sigma=\sigma\vec{v}_0+\vec{\omega}\times\vec{r}
\eeq
for a given spin quantization axis, with $v_0>>\omega r$. We assume furthermore that the vector $\vec{v}_0$ is
  parallel or antiparallel to $\vec{\omega}\times\vec{r}$.  The $difference$ in the centripetal force for a rotating and non-rotating
superconductor is
\beq
\vec{F}_c(\vec{\omega})-\vec{F}_c(0)=-2m_ev_\sigma\omega \hat{r}=\frac{e}{c}\vec{v}_\sigma\times\vec{B}
\eeq
i.e. precisely the Lorentz force provided by the generated London field in the rotating superconductor, Eq. (2).
Eq. (6) implies that the $charge$ $current$ in the rotating superconductor is given by Eq. (4) and in particular is zero in the
absence of body rotation, and that a $spin$ $current$ exists
\beq
\vec{J}_{spin}=\frac{n_s}{2} (\vec{v}_\uparrow-\vec{v}_\downarrow)=n_s \vec{v}_0
\eeq
which is independent of the body rotation and in particular is non-zero when the body is not rotating. The spin current
field Eq. (8) is a function of position.

The existence of such a 'spontaneous' spin current in the absence of body rotation and magnetic field requires the existence of
an electrostatic field $\vec{E}$ in the interior of superconductors, related to $\vec{v}_0$ by Newton's equation
\beq
\frac{d\vec{v}_\sigma}{dt}=\vec{\nabla}\frac{v_0^2}{2}-\vec{v}_0\times(\vec{\nabla}\times\vec{v}_0)=\frac{e}{m_e}\vec{E}
\eeq
Note that the right-hand-side of Eq. (9) is independent of $\sigma$ because the velocity appears squared. 
We have recently proposed a new macroscopic electrodynamic description of superconductors\cite{electro2} which allows for
 the existence of such an
electrostatic field, satisfying the equation
\beq
\nabla^2(\vec{E}-\vec{E}_0)=\frac{1}{\lambda_L^2}(\vec{E}-\vec{E}_0)
\eeq
with $\lambda_L$ the London penetration depth, and $\vec{E}_0$ the electrostatic field originatig in a uniform positive charge density
$\rho_0$ throughout the volume of the superconductor.  

Conversely we may argue that $if$  an electrostatic field exists in the interior of superconductors as predicted by the theory
of hole superconductivity\cite{electro1}, this
 $requires$ the existence of a velocity  field for the supercarriers, and consequently  $requires$ that   each
superfluid carrier have a 'time-reversed' partner of opposite spin moving with opposite velocity so that no charge current results in the
absence of magnetic fields, and hence  leads naturally to the Cooper pair concept as well as to the conclusion that macroscopic
spin currents should exist in the superconducting state.

Next we turn to the microscopic theory. The pairing interaction in the model of hole superconductivity in  a hole representation is given by\cite{hole1}
\beq
V_{k k'}=U+2\alpha (\epsilon_k+\epsilon_{k'})
\eeq
where the 'kinetic interaction' term $\alpha\equiv \Delta t/t$ originates in the correlated hopping term in the Hamiltonian with
amplitude $\Delta t$\cite{hole1}, and where $t$ is the single hole hopping amplitude and $\epsilon_k$ is the hole kinetic energy.

We have argued elsewhere that the model of hole superconductivity predicts that negative charge is expelled from the interior of the
superconductor towards the surface in the superconducting state\cite{electro1}. 
This gives rise to a macroscopic outward pointing electric field in the interior of the
superconductor, and in the presence of such a field the electronic potential no longer has local  inversion symmetry even if the underlying
lattice structure does, hence  the spin-orbit interaction necessarily leads
to splitting of (k,-k) degeneracy. This then leads necessarily to parity breaking and $(u_k,v_k)\neq (u_{-k},v_{-k})$ in the BCS state.

For simplicity we consider a two-dimensional  or quasi-two-dimensional system in a sample of cylindrical symmetry 
and quantize the spin in the perpendicular direction. To allow for the possibility of spin splitting we assume at the outset for the   kinetic energy of a hole of spin $\sigma$ the form
\beq
\epsilon_{k\sigma}=-t\sum_{\nu=x,y} (cos k_\nu +\sigma b sin k_\nu)
\eeq
where we assume unit lattice spacing. The kinetic energy Eq. (12) is time-reversal invariant
but breaks parity for $b\neq 0$. In addition to being induced by charge expulsion, such a 'spin-split' state is favored by a   nearest neighbor off-diagonal matrix element of 
the Coulomb interaction
\beq
H_J=J\sum_{<ij>\sigma}c_{i\sigma}^\dagger c_{j\sigma}  c_{j,-\sigma}^\dagger c_{i,-\sigma} 
\eeq
with $J>0$, 
which exists  in all solids as discussed in \cite{ss}. Of course it can also arise in a crystal without inversion symmetry  through  ordinary spin-orbit coupling.

\begin{figure}
\resizebox{6.5cm}{!}{\includegraphics[width=7cm]{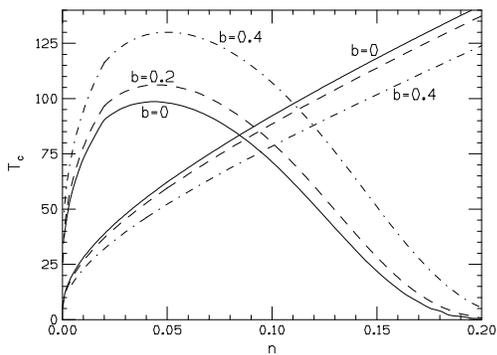}}
  \caption{Critical temperature versus carrier (hole) concentration for the model of hole superconductivity (bell-shaped curves) and for the attractive
  Hubbard model (monotonically increasing curves). The solid, dashed and dot-dashed lines correspond to spin-splitting parameter values
  $b=0$, $b=0.2$ and $b=0.4$ respectively. Parameters in the model of hole superconductivity are $U=5$, $K=1.65$, $t=t_h+n K/8$ with $t_h=0.03$. 
  In the attractive Hubbard model $U=-.15$, K=0, $t=0.04$. Energies are in $eV$ and temperature in $K$. }
   \end{figure}

The kinetic energy in Eq. (11) is then $\epsilon_{k\uparrow}$ or $\epsilon_{-k\downarrow}$ which are identical, and solution to the
BCS equations proceeds identically to the case in the absence of spin splitting, except that in the energy versus $k$ relation the form
Eq. (12) is to be taken. The gap function is of the form\cite{hole1}
\beq
\Delta_k=\Delta_m(-\frac{\epsilon_k}{D/2}+c)
\eeq
with $\Delta_m$, $c$ constants. Hence $\Delta_k\neq \Delta_{-k}$ in general, unless $\Delta_m\rightarrow 0$ , $\Delta_mc\neq 0$
which only occurs for $\alpha\rightarrow 0$, $U<0$ for the interaction Eq. (11).

We show in Figure 2 results for the critical temperature versus hole concentration for a case with parameters
appropriate to the physics of the model of hole superconductivity ($U>0$, $\alpha>0$) compared with an attractive Hubbard
model ($U<0$, $\alpha=0$). 
Notably it is seen in Fig. 2 that spin-splitting enhances the tendency to superconductivity when the
attraction leading to Cooper pairing is of kinetic origin, and suppresses it when the attraction is of potential origin as in the
attractive Hubbard model. It is likely that the same qualitative behavior as in the attractive Hubbard model will result for models
with more extended attractive density-density interactions (eg Hubbard model with attractive nearest neighbor interactions) as well 
as for conventional electron-phonon interaction models (Holstein, SSH, Frohlich) where the effective electron-electron
interaction albeit retarded also leads to potential energy lowering.

In the superconducting state, we find that 
spin splitting raises the condensation energy in the model of hole superconductivity  and lowers
 it in the attractive Hubbard model. Results are shown in Fig. 3 for two values of the hole concentration. We find that for the entire concentration range where
$T_c$ is not zero spin splitting increases the condensation energy in the model of hole superconductivity and decreases it in the attractive
Hubbard model.

\begin{figure}
\resizebox{6.5cm}{!}{\includegraphics[width=7cm]{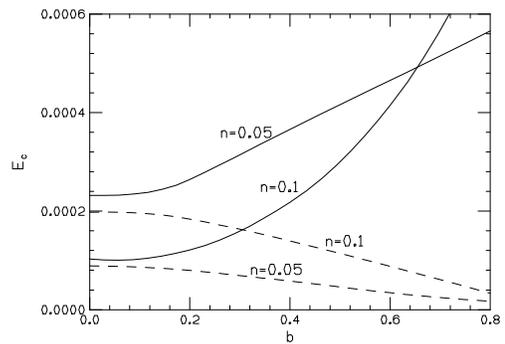}}
\caption{Condensation energy dependence on spin splitting parameter $b$ in the ground state for the model of hole superconductivity (full lines) and
 the attractive Hubbard model (dashed lines) for two values of the hole density. Parameters are the same as in Fig. 2.}
\label{atom4}
\end{figure}

Furthermore in the presence of spin splitting the relation between chemical potential and carrier concentration changes. Figure 4 shows the hole concentration
in the model of hole superconductivity as  a function of spin splitting for fixed value of the chemical potential, showing that spin splitting causes
an $increase$ in the hole concentration for fixed chemical potential. We interpret this result to mean that spin splitting enhances the tendency
for the superconductor to expel electrons from its interior\cite{electro1}. This tendency exists for all hole concentrations.

Of course the model considered here
does not take into account the large charging energy that would result if the hole concentration changed by the amounts shown
in Fig. 4. In fact, as electrons move out of the interior towards the surface the resulting cost in Coulomb energy will stabilize the excess
electron concentration near the surface at a value corresponding to just a few electrons per million atoms, as discussed in ref. \cite{electro1}.

\begin{figure}
\resizebox{6.5cm}{!}{\includegraphics[width=7cm]{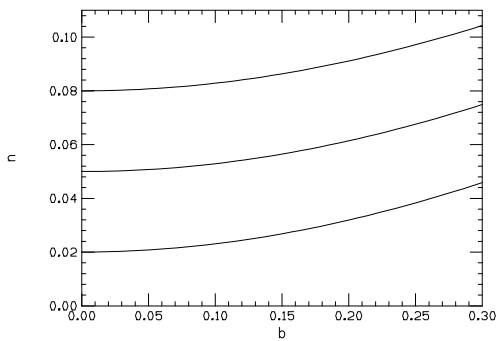}}
\caption{Dependence of carrier concentration on spin splitting parameter for the model of hole superconductivity for fixed value of the
chemical potential. Parameters are the same as in Fig. 2.
}
\label{atom4}
\end{figure}

Furthermore as charge expulsion takes place the resulting internal electric field that builds up further enhances the tendency to spin splitting through the  spin-orbit interaction that occurs between the superfluid electrons in macroscopic orbits and the macroscopic electric field that
builds up, as would occur in a 'giant atom'\cite{atom}.

Finally, our physical picture  provides a simple way to understand how the macroscopic spin current builds up when the system makes
the transition to the superconducting state. As negative charge is expelled from the interior of the superconductor   in the radial direction, 
the interaction of the moving magnetic moment of the electron with
the positive background gives rise to a force in direction perpendicular to the motion whose sign depends on the spin orientation. 
Such an effect was discussed by us in connection with a disipationless intrinsic 
mechanism for the spin Hall effect and the anomalous Hall effect in ferromagnetic metals\cite{ferrohall},
and may also be predicted by a universal intrinsic spin Hall effect recently discussed\cite{univ1,univ2}.
This force   deflects the electron in opposite  azimuthal directions for up and down spin electrons\cite{ferrohall} such as to give rise to the macroscopic spin current. This is shown schematically in Figure 5.

\begin{figure}
\resizebox{6.5cm}{!}{\includegraphics[width=7cm]{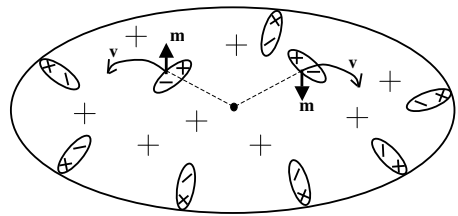}}
\caption{Electrons are expelled towards the surface when the system goes superconducting, and are
deflected by the interaction of the moving magnetic moment (equivalent to an electric dipole moment) with
the positive background, giving rise to a spin current.  The figure also shows the resulting equivalent dipole
configuration near the surface that arises from the macroscopic spin current.
}
\label{atom4}
\end{figure}

In summary we have provided macroscopic and microscopic arguments in favor of the suggestion that macroscopic spin supercurrents exist
in the superconducting state of all metals. As already pointed out by Bohm\cite{bohm}, 'Bloch's theorem' does not invalidate such possibilities.
The direction of flow of these currents is determined by the sample geometry
through spin-orbit coupling to the electrostatic field that results from solution of the macroscopic electrodynamic equations\cite{electro2}. Spin currents
further stabilize the superconducting state in the model of hole superconductivity and explain the value of the 'London field' in rotating
superconductors. When an external magnetic field is applied the pre-existent spin currents will be modified so that they no longer
exactly cancel giving rise to a Meissner current that screens the applied magnetic field. Experimental detection of
macroscopic spin currents in superconductors may be possible through spin-resolved neutron scattering experiments
as discussed in ref. \cite{ss} or angle resolved photoemission experiments with circularly polarized photons as proposed in ref. \cite{varma}
or optical \cite{driel} or other experiments\cite{atom}. It has also been suggested that the physics discussed here could have relevance to aromatic molecules\cite{arom1,arom2}.


\begin{references}
\bibitem{clarify} Note that changing $e$ to $2e$ does not solve this problem because $m_e$ would have to change to $2m_e$, leaving Eqs. (2) and (5) unchanged.
\end{references}
\end{document}